\documentclass[aa_abbrev,aas_macros,useAMS,usenatbib]{mn2e}

\usepackage{amssymb,amsmath}
\usepackage{graphicx}		%! Preamble.
\usepackage{import}           	%! needed to include files in other dirs
\usepackage{rotating}		%! needed to rotate images and tables
\usepackage{array,multirow}   	%! needed to enhance tabular 
\usepackage{txfonts}		%! needed for different fonts
\usepackage{defn}		%! my own definitions

\title[X-ray observations of \suuma{}]
{X-ray observations of \suuma{} throughout six outbursts}

\author[D. J. Collins and P. J. Wheatley]
    {David J. Collins\thanks{E-mail: D.J.Collins@warwick.ac.uk} and 
    Peter J. Wheatley\thanks{E-mail: P.J.Wheatley@warwick.ac.uk} \\
    Department of Physics, 
    University of Warwick, 
    Coventry CV4 7AL, UK}

\begin{document}

	\date{\today}
	\pagerange{\pageref{firstpage}--\pageref{lastpage}} \pubyear{2009}
	\maketitle
	\label{firstpage}

	\graphicspath{{images/}}

\begin{abstract}

We present \rxte{} observations covering six normal outbursts of the
dwarf nova \suuma{}, the prototype of its class.  The outbursts showed
consistent X-ray behaviour with the X-ray count rate dropping suddenly
by a factor of four, and with evidence for a half day delay between
the optical rise and the X-ray suppression.  In contrast to \sscyg{},
an X-ray flux increase at the beginning of outburst was not observed,
although it is expected from boundary layer models.  The X-ray flux
was high and decreasing during quiescence, in conflict with the disc
instability model.  The X-ray spectrum of \suuma{} was softer in
outburst than during quiescence, and it was consistent with constant
reflection.

\end{abstract}


\section{Introduction}

% In the soft x-ray band, SU UMa is one of the most luminous dwarf nova
% \citep{1997A&A...327..602V}.  - pg 610(9)

%% Intro to CVs
Cataclysmic variables are close binary systems in which a red dwarf
secondary star transfers mass via Roche lobe overflow to a white dwarf
primary (see \cite{1995cvs..book.....W} for a review).  Material is
accreted by the white dwarf through an accretion disc and in dwarf novae
the disc is unstable and undergoes regular outbursts.  \suuma{}-types
are a subclass of dwarf novae which exhibit two types of outbursts:
normal outbursts and superoutbursts.  Superoutbursts occur less
frequently, last longer and are brighter than normal outbursts.

In the standard model about half of the gravitational energy of the
accreting matter is released in a boundary layer between the accretion
disc and the white dwarf \citep{1973A&A....24..337S,
1977MNRAS.178..195P, 1981AcA....31..267T}.  This energy is thought to
be emitted as X-rays due to shock heating as the accreting material
settles on to the white dwarf surface.  Therefore X-ray observations
are sensitive to the mass transfer rate through the accretion disc and
to conditions in the inner accretion disc.

Dwarf novae are sources of hard X-rays in quiescence and observations
of eclipsing systems support the boundary layer model demonstrating
that the X-rays arise in the immediate vicinity of the white dwarf, at
least in quiescence \citep[e.g.][]{1997ApJ...475..812M,
2003MNRAS.345.1009W}.

The X-ray spectrum in quiescence is a hot optically thin plasma
\citep[e.g.][]{1991ApJ...382..290E}.  Observed lines indicate that the
X-ray emitting plasma covers a wide range of temperatures
\citep[e.g.][]{2005MNRAS.357..626B} and is consistent with a cooling
plasma settling onto the white dwarf through a disc boundary layer
\citep[e.g.][]{1996A&A...307..137W, 2005ApJ...626..396P}.

In outburst the hard X-rays are usually suppressed and replaced with
an intense extreme ultraviolet component
\citep[e.g.][]{2003MNRAS.345...49W}. This is thought to be due to the
boundary layer becoming optically thick to its own emission as the
accretion rate increases during outburst \citep{1979MNRAS.187..777P,
1985ApJ...292..550P}. However, the X-ray flux evolution through the
outburst cycle is not well reproduced by the standard disc instability
model \citep[e.g.][]{2001NewAR..45..449L}.

%% summary of su uma observations
%% 	observed $19$ times, $7$ - outburst, $12$ - quiescence
%% 	exposure from $2 - 10$ to $49$ ks 
Most X-ray observations are too short to follow the flux evolution
through outburst, and our observational picture is still based on just
a handful of well studied outbursts:
\sscyg{} \citep{1979MNRAS.186..233R, 1992MNRAS.257..633J, 2003MNRAS.345...49W},
\vwhyi{} \citep{1987MNRAS.225...73P, 1996A&A...307..137W},
\yzcnc{} \citep{1999A&A...346..146V},
\wzsge{} \citep{2005ASPC..330..257W}.
In some cases systems have been seen to deviate from the standard
picture (e.g. \ugem{}, \citealp{1984MNRAS.206..879C}; \gwlib{},
\citealp{2009arXiv0907.2659B}), but it is not clear whether individual
systems exhibit a range of behaviour or whether each system exhibits
consistent behaviour.

In this paper we present X-ray observations spanning six outbursts of
the dwarf nova \suuma{}.  Snapshot observations have shown that in
quiescence \suuma{} is a source of hard X-rays, emitted from a hot
optically thin boundary layer region \citep{1991ApJ...382..290E}. In
outburst the count rate has been seen to drop by a factor of $3$
\citep{1994ApJ...425..829S}.  High resolution data taken with the
\chandra{} \hetg{} during quiescence showed the presence of a weak
fluorescent iron line indicating there may be reflection present in
the system \citep{2006ApJ...642.1042R}.

By monitoring a system throughout multiple outbursts, we aim to
determine whether the X-ray flux evolution is consistent between
different outburst cycles.

\section{Observations}
\label{sec_obs}

\suuma{} was monitored using the \rxte{} Proportional Counter Array
(\pca{}) beginning on $2001$ March $25$ and ended on $2001$ June $21$.
The observations spanned six normal outbursts.  The total exposure was
$336$ ks observed over $193$ visits taken in blocks of $1-14$ ks
exposure with gaps due to Earth occultations and passage of the
spacecraft through the South Atlantic Anomaly (SAA).  All \rxte{} count
rates in this paper are for three \pcu{}s.
% Data is taken from the proportional counter array (\pca{})
% experiment on \rxte{}. It consists of $5$ identical xenon-filled
% proportional counter units (\pcu{}) sensitive to X-rays in the
% energy range $2-60$ keV. All \rxte{} count rates are presented for
% three \pcu{}s.

The data were extracted using \ftools{}, version $6.1.2$, provided by
the High Energy Astrophysics Science Archive Research
Centre.\footnote{http://heasarc.gsfc.nasa.gov/ftools} Only the top
xenon layer was used and the \pca{} data were extracted in
Standard-$2$ binned mode. \saextrct{}, version $4.2$e, was used to
produce light curves with a time resolution of $16$ seconds, and
spectral files binned into $129$ channels. Data were excluded where
the elevation was less than $10$ degrees, the ELECTRON2 ratio was
greater than $0.1$ and the time since SAA passage was under $10$
minutes.  On $2000$ May $13$ \pcu{}$0$ suffered a loss of propane from
its veto layer resulting in a higher number of false events being
observed in that detector. As a result this detector was not used in
the extraction of the data set.  \pcabackest{}, version $3.0$ and
\pcarsp{}, version $10.1$, were used to calculate the background, and
response matrix. Data from different detector configurations were
extracted as separate spectra an re-binned using \rbnpha{}, version
$2.1.0$.

\section{Time series analysis}
\label{sec_time_series}

%%% image created with lc2ps/plotlc_all_op, time=[2450000,2453250]
\begin{figure*}

\center{\includegraphics[width=0.95\linewidth]{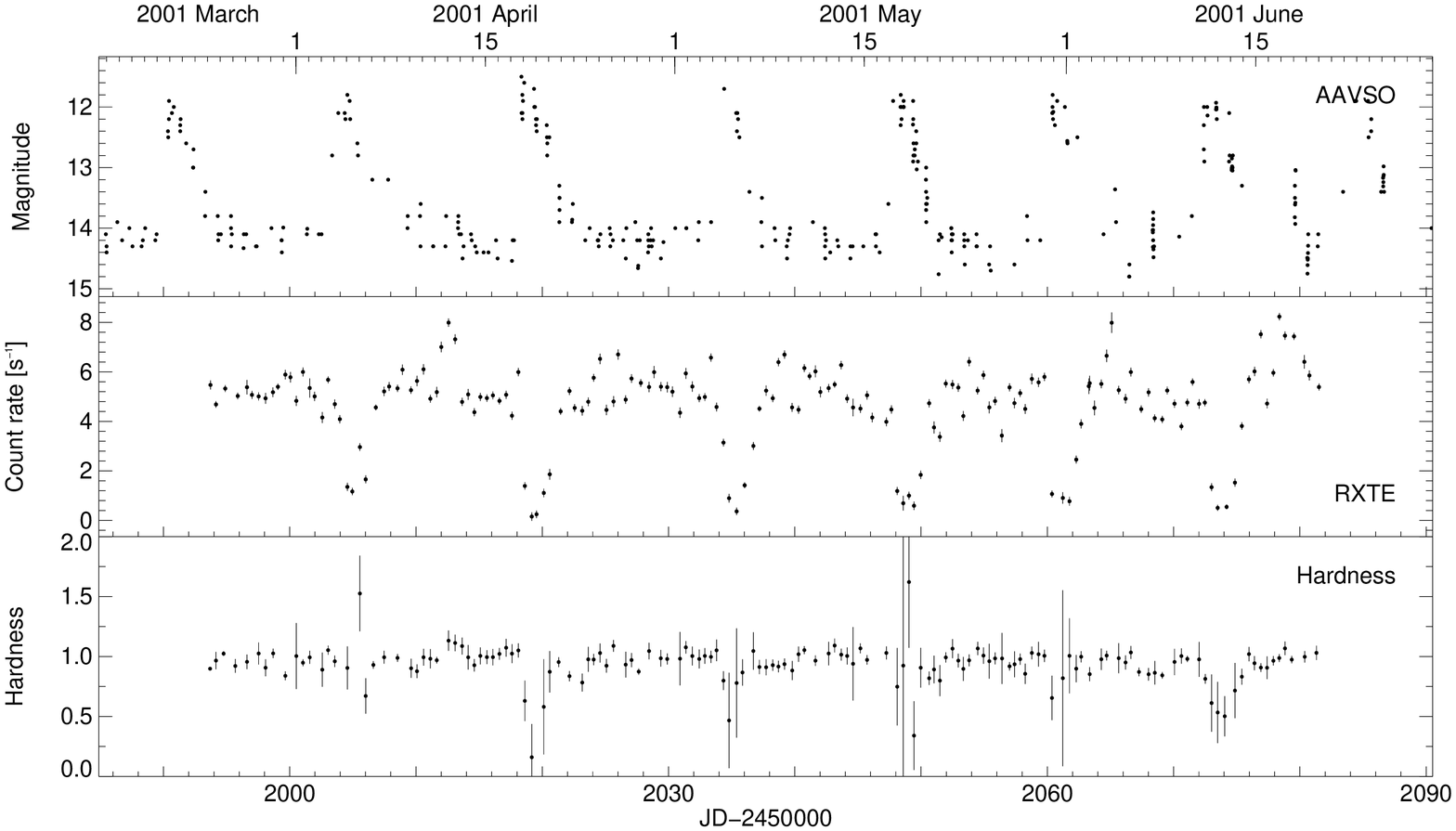}}
\vspace{0.1cm}
\caption{\label{pl_xophard_suuma} \aavso{} (upper panel) and \rxte{}
	(middle panel) observations of \suuma{} during the March - June
	$2001$.  The X-ray hardness ratio, defined as the hard band,
	$3.7-18.5$ keV, divided by the soft band, $2.0-3.7$ keV, is also
	plotted.  The X-ray count rates are for $3$ \pcu{}. }

\end{figure*}

The \rxte{} observations cover six normal outbursts of \suuma{}.  Figure
\ref{pl_xophard_suuma} shows the X-ray observations plotted with optical
observations from the \aavso{}. The optical data consist of visual
observations ($85$ per cent) and V band CCD observations ($15$ per cent)
with an average cadence of $5$ hours (but usually concentrated into
night time in the USA).  In outburst the optical band rapidly brightened
from approximately $14^{th}$ to $12^{th}$ mag in approximately $0.5$
days before declining back into quiescence.  Outbursts were nearly
equally spaced, lasting approximately $2$ days with the quiescence
periods lasting approximately $10$ days.

\begin{figure}

  \center{\includegraphics[width=0.9\linewidth]{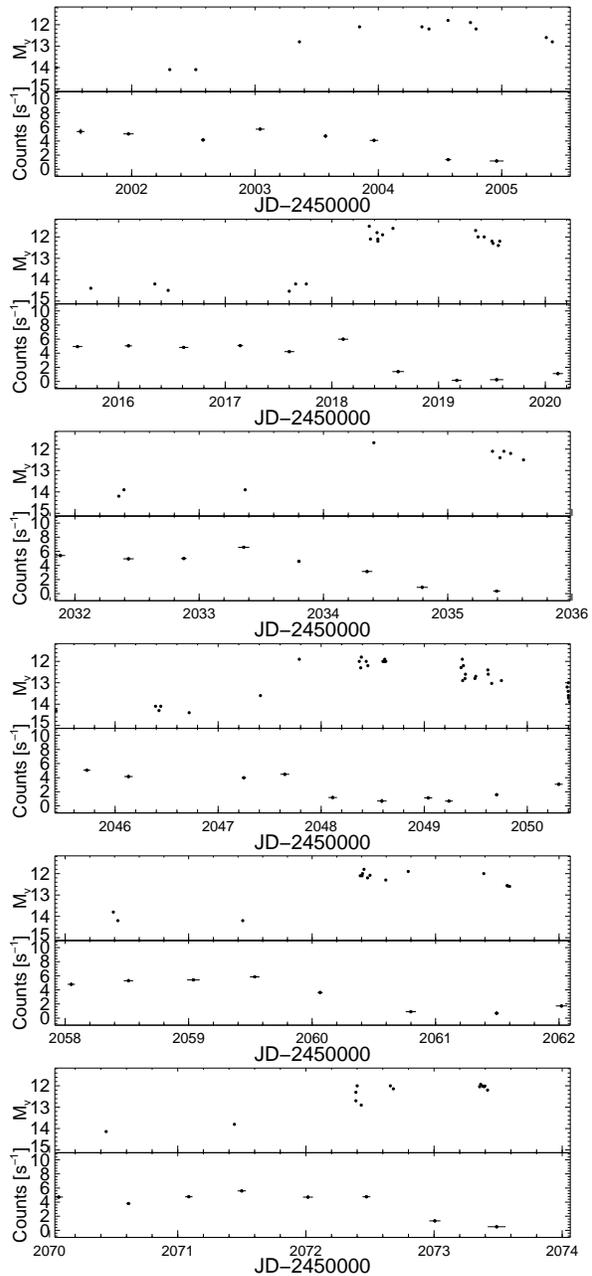}}
  \vspace{0.1cm}
  \caption{\label{pl_ob_start} Detailed sections of the optical (upper
	  panel) and X-ray (bottom panel) light curves from Fig.
	  \ref{pl_xophard_suuma} during the transition to outburst. The
	  X-ray count rates are for $3$ \pcu{}.}

\end{figure}

In the hard X-ray band as the system enters outburst the X-rays become
suppressed to near zero. The X-ray flux remains suppressed for most of
the duration of the outburst, recovering during optical decline.  The
beginning of each outburst is shown in more detail in Fig.
\ref{pl_ob_start}, and there is no sign of a peak preceeding the X-ray
suppression, as was observed in SS Cygni. The cadence of X-ray
observations at the beginning of the outbursts was typically $12$ hours.

Also plotted in Fig. \ref{pl_xophard_suuma} is the X-ray hardness
ratio which was created using counts in the energy bands $2.0-3.7$ keV
and $3.7-18.5$ keV. During quiescence the hardness ratio does not show
any obvious variations, although at JD $2452012$ for approximately $4$
days a higher X-ray count rate was accompanied by a harder spectrum.
%Calculating the $\chi^2$ during this period, using the average of the
%hardness ratio as the model hardness, results in the $\chi^2$ of 
%$4.2$ ($3$ d.o.f) with a probability of $0.76$ suggesting that this 
%feature is significant.  
Despite large error bars during outburst it is clear that the system is
softer during outburst than during quiescence . This has also been seen
in other dwarf novae \cite[e.g.][]{2005MNRAS.357..626B}. Closer
inspection of Fig.  \ref{pl_xophard_suuma} shows that the softening of
the spectrum coincides with the hard X-rays becoming quenched.

%%% image created with lc2ps/plotlc_obhr

\begin{figure}

  \center{\includegraphics[width=0.9\linewidth]{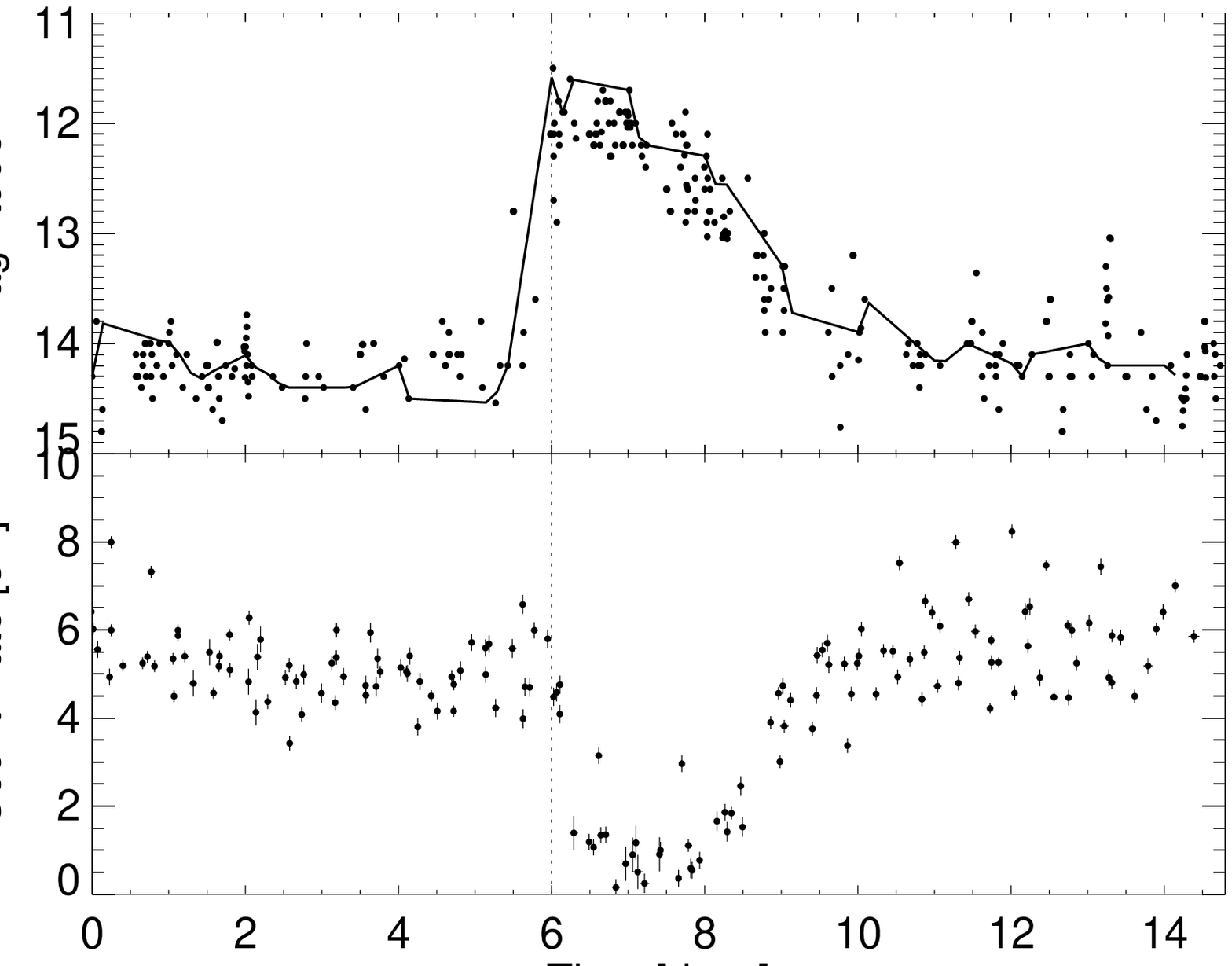}}
  \vspace{0.1cm}
  \caption{\label{pl_xop_corr_suuma} The \aavso{} (upper panel) and
      \rxte{} (bottom panel) outbursts from Fig.
      \ref{pl_xophard_suuma} auto correlated and overlaid. The second
	  optical outburst (thick line) was used as the template. The X-ray
	  count rates are for $3$ \pcu{}.}

\end{figure}

To make comparison between the outbursts a composite light curve was
made of the six outbursts, Fig. \ref{pl_xop_corr_suuma}.  The times were
shifted to maximise the auto-correlation in the optical band using the
second outburst as a template (JD $2452017$). The composite light
curve shows that the shapes of the six outbursts were highly
repeatable.  In both the optical and X-ray bands the quiescent periods
following outburst were more variable than prior to the outbursts.  A
peak in the count rate immediately before outburst was also not
observed in the composite light curve, which has an average separation
of points of only $2$ hours.   The X-ray suppression is remarkably
rapid and is not resolved in the composite light curve, suggesting it
occurs faster than two hours.

Figure \ref{pl_xop_corr_suuma} also shows there is a delay between the
start of outburst in the optical and X-ray bands, however, due to the
sparse sampling of both optical and X-ray data it is difficult to
determine the precise duration of the delay.  To quantify the delay the
data were interpolated and times when the X-ray and optical flux crossed
the mid point between outburst and quiescence were calculated, these
times are shown in Table \ref{tab_suuma_delay}.  The delays show that
the X-ray suppression follows the optical rise by $0.25-1.12$ days
later, with a median delay of $0.57$ days.  The range in delay times can
be accounted for by the cadence of the observations (also listed in
Table \ref{tab_suuma_delay}).

%%% image created with lc2ps/plotlc_overlay, template=2, optical=0

% insert delay table 
\begin{table}
  \begin{minipage}[c]{\columnwidth}

	\caption{\label{tab_suuma_delay}Optical and X-ray times for the flux
		to reach mid-point of the transition to outburst, including the
		cadence of the optical and X-ray observations respectively.  }

%%
%% table with information about optical-outburst delay
%% table includes:
%%		optical time
%%		X-ray time
%%		delay
%%

%\begin{tabular*}{\textwidth}{p{3cm}@{\extracolsep{\fill}}cccc}
\begin{tabular*}{\textwidth}{c@{\extracolsep{\fill}}cccc}

\hline \hline

Optical time &
Cadence &
X-ray time &
Cadence &
Delay \\

%% units
[JD] &
[days] &
[JD] &
[days] &
[days] \\

\hline

%mid outburst        3.3794543       13.355730
%average delay       0.61751243

2452003.0       &0.79  &2452004.1     &0.60  &1.12 \\
2452017.9       &0.47  &2452018.4     &0.46  &0.45 \\ 
2452033.6       &1.03  &2452034.3     &0.42  &0.68 \\ 
2452047.5       &0.39  &2452047.8     &0.42  &0.34 \\ 
2452059.8       &0.88  &2452060.1     &0.68  &0.25 \\
2452071.8       &0.88  &2452072.7     &0.47  &0.86 \\

\hline

\end{tabular*}

  \end{minipage}
\end{table}

\begin{figure}

  \center{\includegraphics[width=0.9\linewidth]{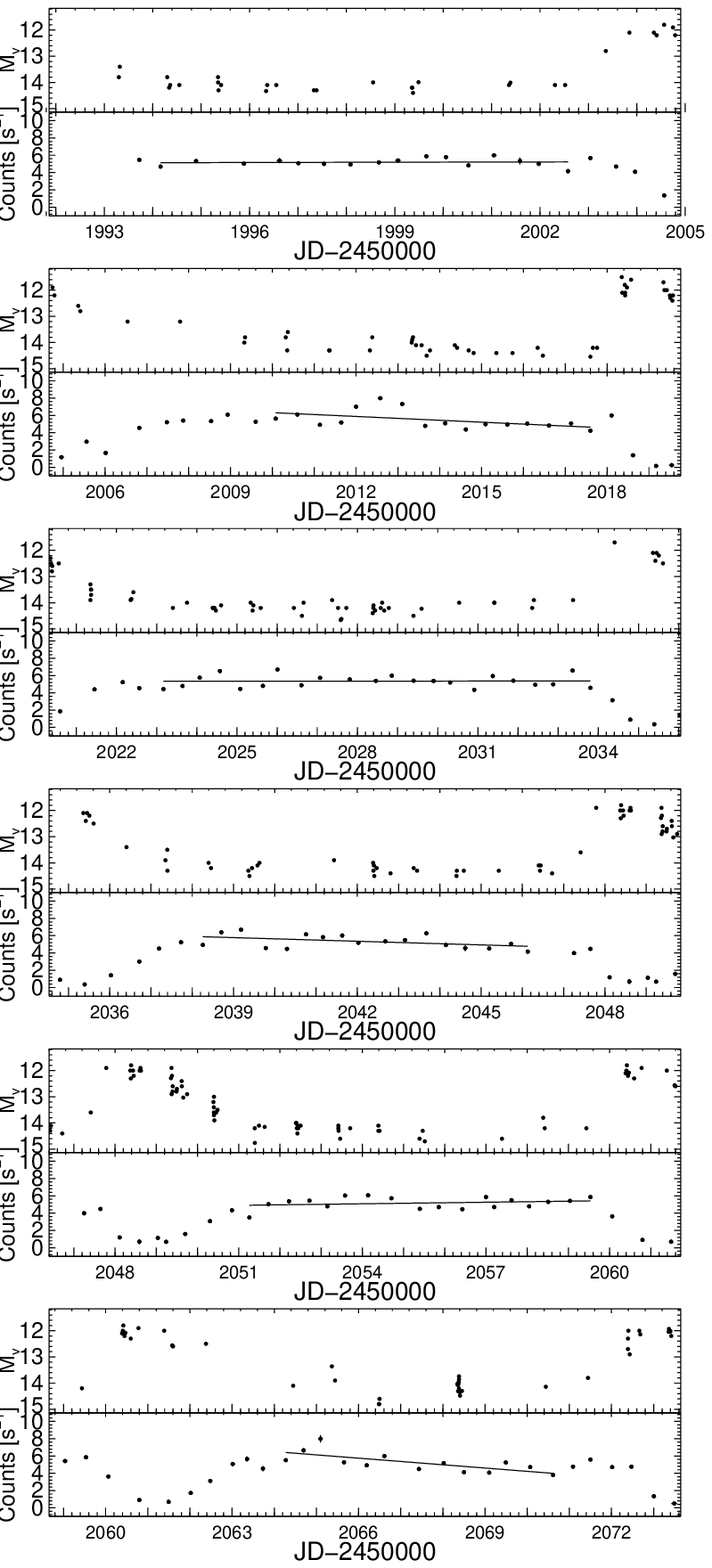}}
  \vspace{0.1cm}
  \caption{\label{pl_q_declinei} Individual optical (upper panel) and
	  X-ray (bottom panel) light curves from Fig.
	  \ref{pl_xophard_suuma} centred around the quiescent intervals.
	  Each quiescent X-ray interval is also plotted with a linear least
	  squares fit at times when the optical magnitude was fainter than
	  $14^{th}$ mag. The X-ray count rates are for $3$ \pcu{}. }

\end{figure}

\begin{figure}

\center{\includegraphics[width=0.9\linewidth]{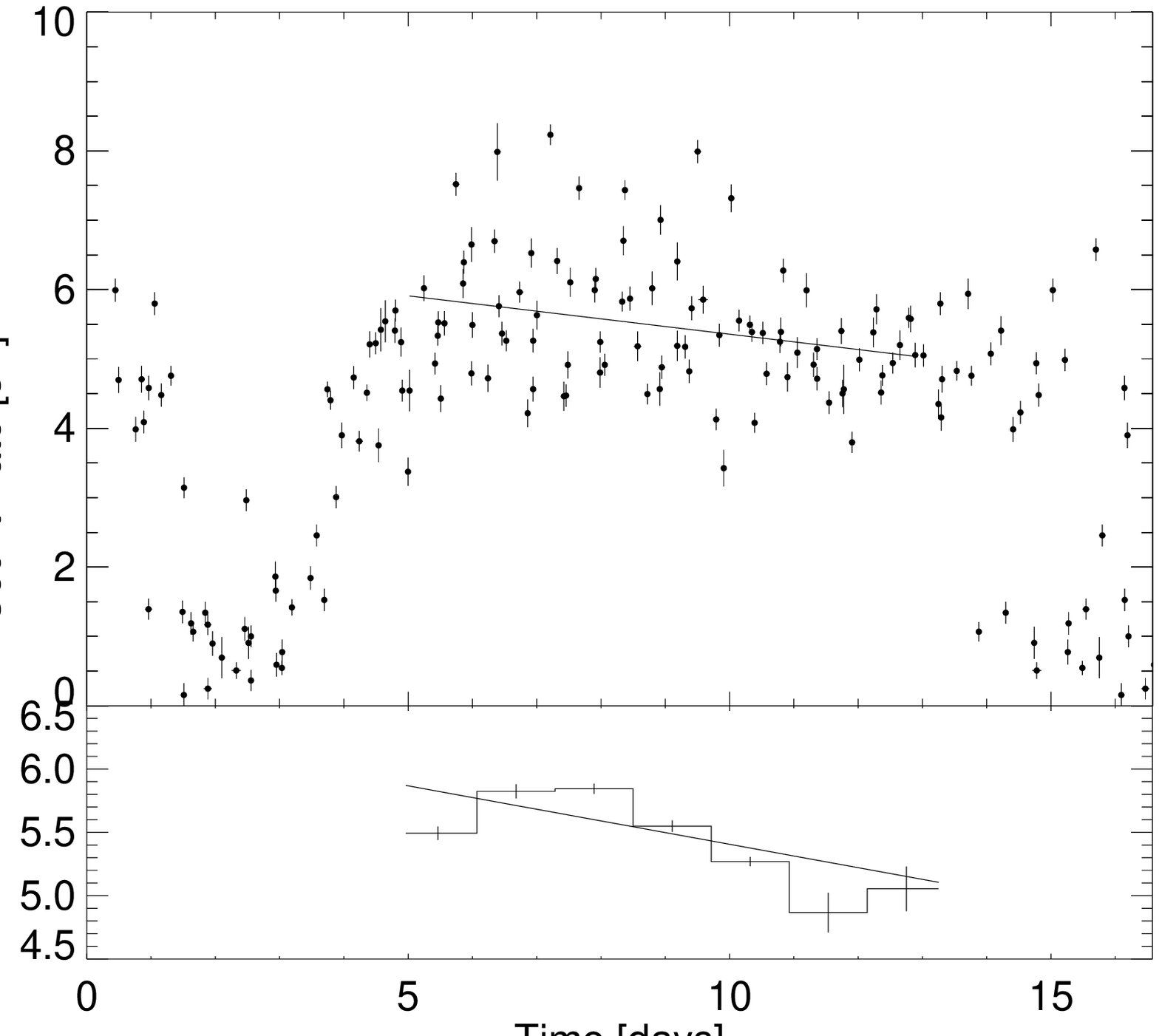}}
\vspace{0.1cm}
\caption{\label{pl_q_decline} Composite X-ray light curve from Fig.
	\ref{pl_xop_corr_suuma} centred around the quiescent interval with a
	linear least squares fit to the quiescent data (top panel) and to
	the binned quiescent data (bottom panel) at times when the optical
	magnitude was fainter than $14^{th}$ mag. The X-ray count rates
	are for $3$ \pcu{}.}

\end{figure}

In order to determine whether the X-ray flux increases during
quiescence, as predicted by the disc instability model, we performed a
linear least squares fit to the quiescent intervals, plotted in Fig.
\ref{pl_q_declinei}.  The X-ray observations were selected when the
optical light curve was fainter than $14^{th}$ mag.  There is short
timescale variability in the light curves, but the quiescent trend is
well represented by the best fit lines.  The fits show that the
quiescent X-ray count rate either remains constant or decreases.  The
gradients of the line fits are presented in Table \ref{tab_q_decline}.
Figure \ref{pl_q_decline} shows the composite light curve of Fig.
\ref{pl_q_declinei} with a linear least squares fit to the all the
quiescent data. The fit was applied to the interval where the optical
was fainter than $14^{th}$ mag in all six cycles.  It shows that, in
addition to short timescale variability, there is an overall decline
in the hard X-ray flux during quiescence.  The mean count rate
decreased from a maximum of $5.8$ counts s$^{-1}$, with a gradient of
$0.09$ counts s$^{-1}$ day$^{-1}$, although the average decline is not
well represented by a straight line ($\chi^{2}_{\nu} = 4.9$ with $5$
degrees of freedom).  A similar decline during quiescence has
previously been seen in \sscyg{} by \cite{2004ApJ...601.1100M}.

%%% image created with lc2ps/plotlc_obhr

% insert decline table 
\begin{table}
  \begin{minipage}[c]{\columnwidth}

	\caption{\label{tab_q_decline} Gradients for the linear least
		squares fits to the quiescent X-ray data from Fig
		\ref{pl_q_declinei} and Fig \ref{pl_q_decline}. }

%%
%% table with information about quiescent decline
%% table includes:
%%		quiescent time
%%		decline
%%

%\begin{tabular*}{\textwidth}{p{3cm}@{\extracolsep{\fill}}cccc}
\begin{tabular*}{\textwidth}{c@{\extracolsep{\fill}}c}

\hline \hline

Quiescence interval&
Gradient \\

%% units
[JD] &
[counts s$^{-1}$ days$^{-1}$] \\

\hline

2451999  	&$+0.01 \pm 0.02$ \\
2452013     &$-0.22 \pm 0.02$ \\
2452028     &$+0.00 \pm 0.01$ \\
2452042     &$-0.14 \pm 0.02$ \\
2452055     &$+0.06 \pm 0.02$ \\
2452068     &$-0.38 \pm 0.03$ \\

%Composite	&$-0.11 \pm 0.01$ \\
Composite	&$-0.09 \pm 0.02$ \\

\hline

\end{tabular*}

  \end{minipage}
\end{table}

%%% image created with lc2ps/plotlc_all_op, time=[2450000,2453250]

\section{Spectral fitting}
\label{sec_spec_fitting}

\subsection{Combined quiescence spectrum}
\label{sec_q}

The data were initially binned into one quiescent spectrum containing
data with count rates higher than $3$ counts$^{-1}$ (per $3$ \pcu{}s),
resulting in a total exposure of $251$ ks.  We adopted a systematic
error of $0.5$ per cent \cite[e.g.][]{1999ApJ...522..460W,
2006ApJS..163..401J, 2006A&A...447..245W} applied to the background
subtracted data.  Since \suuma{} is faint the systematic uncertainty
in the background is likely to contribute significantly, this will be
addressed in more detail in Section \ref{sec_back_syst}.  The data
were initially fitted with the error on the background set to zero.

\begin{figure}

\center{\includegraphics[width=0.9\linewidth]{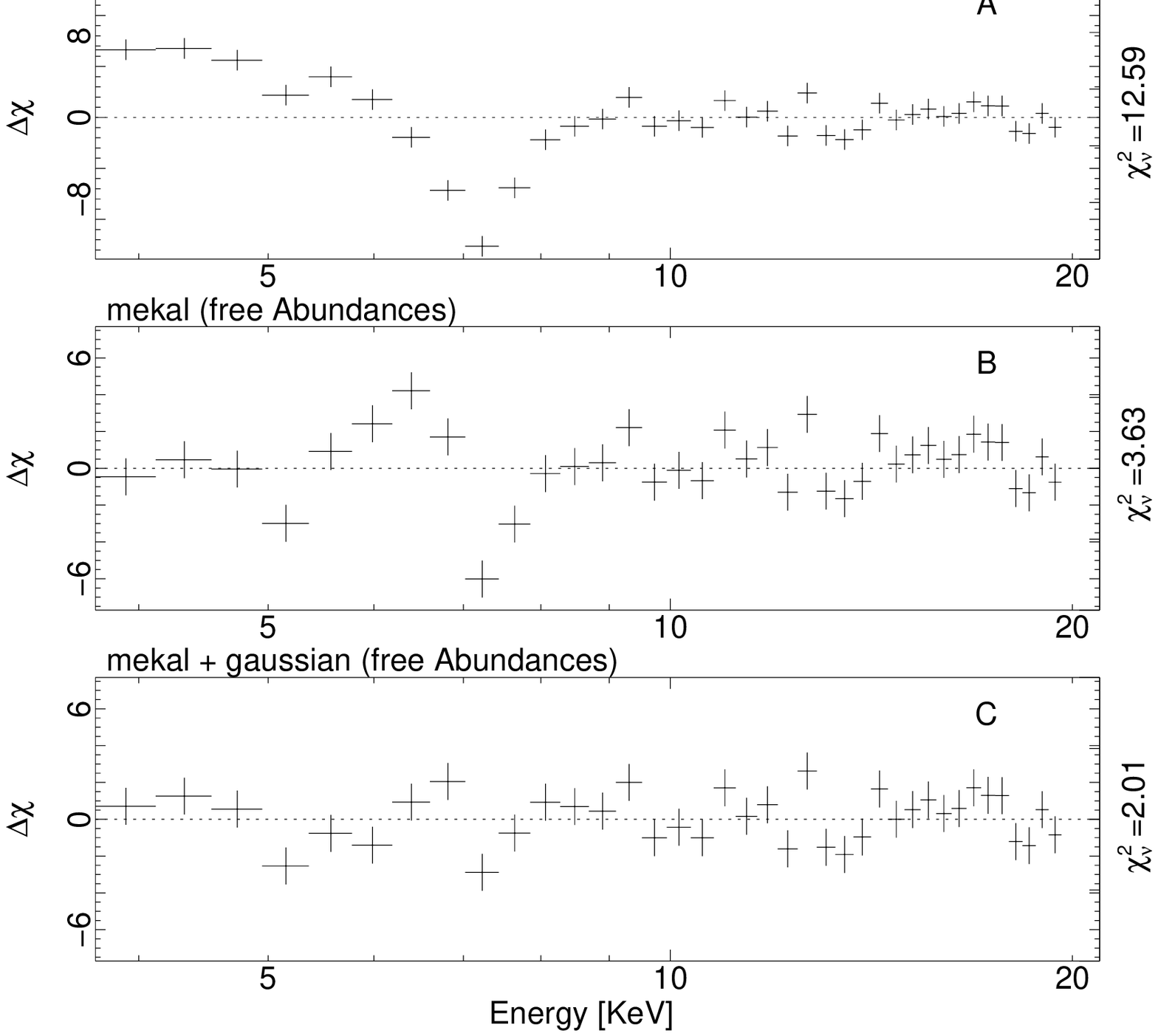}}
\vspace{0.1cm}
\caption{\label{pl_2spec_q} The top panel shows the data (crosses) and
	the folded model of a single temperature continuum model for
	the combined quiescence spectra (Section \ref{sec_q}).  The
	panels below show the residuals for this model and fits to
	subsequent models. The model, as used in \xspec{}, is labelled
	above the residuals, with the $\chi^{2}_{\nu}$ at the end of the
	individual panel.  All errors are in the $68$ per cent confidence
	interval for one parameter of interest ($\Delta\chi^{2} = 1.0$) }

\end{figure}

A single temperature thermal plasma model was fitted to the data; the
mekal model in \xspec{} \citep{1985A&AS...62..197M,
1986A&AS...65..511M, 1995ApJ...438L.115L}.  Applying this model to the
quiescent spectrum did not result in a good fit with a
$\chi^{2}_{\nu}$ of $13$ ($36$ d.o.f). The residuals, Fig.
\ref{pl_2spec_q} panel A, show that there is an excess in the model
between $6-8$ keV. Comparing the plotted model and data it is apparent
that this excess comes from the modelled iron line emission in the
thermal plasma model. Reducing the strength of the thermal lines can
be achieved by lowering the metal abundances.  Making the model
abundances a free parameter gave a significant improvement to the fit
resulting in a $\chi^{2}_{\nu}$ of $3.6$ ($35$ d.o.f) with an ftest
probability of $2 \times 10^{-11}$.  The residuals to this model can
be seen in Fig. \ref{pl_2spec_q} panel B.  The best fit abundances
were found to be lower than solar levels \citep{1989GeCoA..53..197A}
at $0.64 \pm 0.01$ solar.  Sub-solar abundances have been found for
other dwarf novae (\citealp[e.g.][]{2005MNRAS.357..626B}). 
% IMPLICATIONS OF LOWER A probably due to slower evolution of lower
% mass stars? or maybe different progenitors if metalicities are
% lower.

The modification to the model reduced the size of the excess in the
residuals (Fig. \ref{pl_2spec_q} panel B) but it was still poorly
fitting in the $6-8$ keV energy range. The residuals show a deficit in
the model between $6-7$ keV providing evidence for a $6.4$ keV
fluorescence line of neutral iron, also seen in \suuma{} by
\cite{2006ApJ...642.1042R}.  A narrow Gaussian component fixed at
$6.4$ keV was added to the model, and the result of this addition is
seen in Fig. \ref{pl_2spec_q} panel C.  The fit produces a
$\chi^{2}_{\nu}$ of $2.01$ ($34$ d.o.f). The addition of the narrow
emission line improved the fit between $6-8$ keV, but it was unable to
remove the residual feature near $7$ keV.

%%  model plots were made with background=0 systematics=0.5
%%% image created with lc2ps/plotlc_all_op, time=[2450000,2453250]

%
%% background systematic error
%
\subsection{Background systematics}
\label{sec_back_syst}

The unacceptably high $\chi^{2}_{\nu}$ produced by these fits
indicates that either the model is inadequate, or systematic
uncertainties are present in the data that are not fit by the model.
Extensive analysis of the calibration and background model for the
\pca{} was carried out by \cite{2006ApJS..163..401J}.  They determined
that the energy calibration has deviations of $\leq 1$ per cent from
power law fits to the Crab Nebula and unmodeled variations in the
instrumental background at $\leq 2$ per cent below $10$ keV and $\leq
1$ per cent between $10-20$ keV.  When \saextrct{} generates the data
files the error is calculated as the square root of the number of
counts in the data file. This is correct for the data, however, the
background is estimated from the average of a huge amount of data and
so the resulting error on the background is
overestimated.\footnote{http://astrophysics.gsfc.nasa.gov/xrays/programs/rxte/pca/chisquare.html}
A value of zero was used as the background error in the fitting in
Section \ref{sec_q}. However, \suuma{} is a faint system, so
background systematics are likely to contribute significantly.

\begin{figure} 
  \center{\includegraphics[width=0.9\columnwidth]{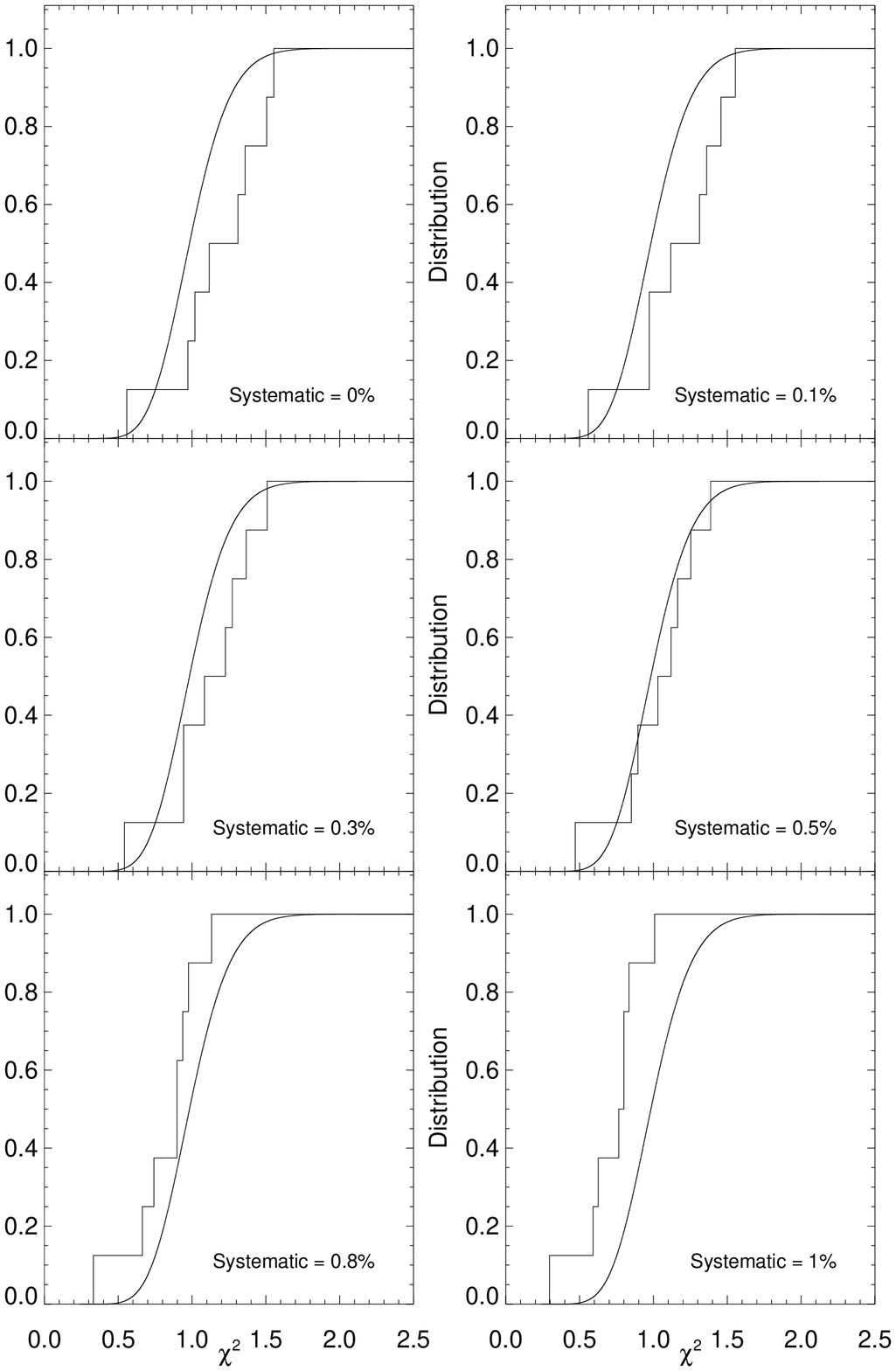}}

  \caption{\label{pl_s2spec}Cumulative histogram plots of
    $\chi^{2}_{\nu}$ resulting from fits of all spectra to the thermal
    plasma model with a narrow emission line. The foreground
    systematic error is fixed at $0.5$ for all fits while the
    background systematic is varied from $0$ to $1$ per cent. }

\end{figure}

%
%% use 8 spectra
%
To investigate the level of systematic error required on the
background, the data were re-binned forming eight spectra.  Using the
same criteria as above, observations with count rates lower than $3$
counts$^{-1}$ (per $3$ \pcu{}s), based on Fig.  \ref{pl_xophard_suuma},
were considered to be in outburst and were binned into one outburst
spectrum containing $48$ ks of exposure.  One spectrum was made for each
quiescent period between the outbursts with each quiescent spectrum
containing an average of $36$ ks of exposure.  A systematic error of
$0.5$ per cent was applied to the background subtracted spectrum as
before, and a histogram of the $\chi^{2}_{\nu}$ distribution was
produced from the model fits.  A series of fits were made with the
systematic error on the background selected in the range $0-1$ per cent,
as identified by \cite{2006ApJS..163..401J}. The data were fitted using
the current best model: a thermal plasma model with free abundances and
a narrow emission line fixed at $6.4$ keV. Figure \ref{pl_s2spec} shows
the cumulative histogram plots of the best fit $\chi^{2}_{\nu}$
distributions, with the cumulative plot of the $\chi^{2}$ distribution
(curve) on the same plot.  The plots show that a systematic error of $0$
and $0.1$ per cent in the background underestimates the error resulting
in distributions that are greater than $1$. Similarly a systematic error
of between $0.8$ and $1$ per cent overestimate the error and result in
distributions that are less than $1$. A systematic error between $0.3$
and $0.5$ per cent produces a distribution that is closest to a
$\chi^{2}$ distribution, with a background systematic error of $0.5$ per
cent producing the best result.  Since this is within the range
identified by \cite{2006ApJS..163..401J} and produced acceptable fits to
the model, this value was adopted and applied to all data.

Fitting the quiescent and outburst spectra with the new background
systematic error resulted in an acceptable $\chi^{2}_{\nu}$ for all
spectra. We also found an acceptable fit with a $\chi^{2}_{\nu}$ of
$0.99$ ($34$ d.o.f) to the combined quiescent spectrum in Section
\ref{sec_q}.

%
%% fitting all spectra
%
\subsection{Time resolved spectra}
\label{sec_res_spec}

\begin{figure}

\center{\includegraphics[width=\linewidth]{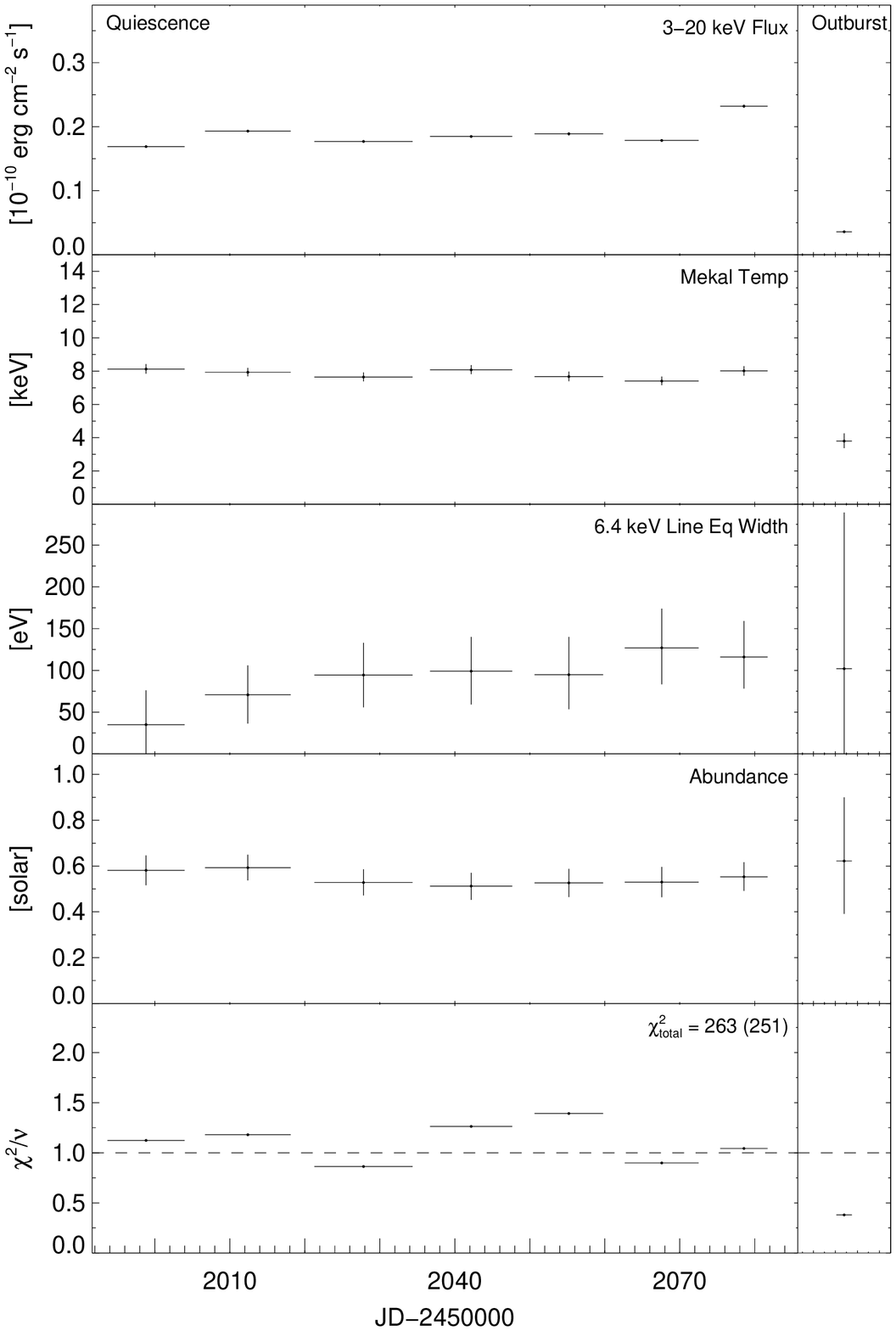}}
\caption{\label{pl_7spec_suuma} Thermal plasma model with the addition
	of a fixed $6.4$ keV emission line component due to fluorescence.
	X-axis error bars mark the start of the first observation until
	the end of the last observation of the original spectra used to
	produce that spectrum. The spectra to the far right of the plot
	contains all outburst data and is thus plotted separately. All
	errors are in the $68$ per cent confidence interval for one
	parameter of interest ($\Delta \chi^2 = 1.0$). }

\end{figure}

The best fit parameters resulting from the fit to the thermal plasma
model with narrow emission line are shown in Fig.
\ref{pl_7spec_suuma}.  The flux of the quiescent spectra varies during
the observation with the seventh spectrum emitting more X-rays on
average than the other quiescent spectra. The system is also fainter
and softer in outburst than in quiescence. Within error the
temperature of the quiescent spectra are consistent with each other at
an average temperature of $7.8 \pm 0.3$ keV. The outburst spectrum
fitted a lower temperature of $3.8 \pm 0.4$ keV. The best fit free
abundances were also consistent with each other producing an average
abundance of $0.64 \pm 0.01$ solar. The outburst spectrum was less
well constrained fitting an abundance of $0.62 \pm 0.25$, however it
is still consistent with the quiescent spectra. The $6.4$ keV line
strength is weaker in the earlier quiescent intervals, but within the
error all spectra are consistent with each other, with an average
equivalent width of $91 \pm 35$ eV during quiescence. The best fit
line strength during outburst is consistent with the quiescent
spectrum but is poorly constrained ($102 \pm 102$ eV).  These
equivalent widths are also consistent with equivalent widths expected
from a semi-infinite, plane parallel cold slab irradiated by an
external source of X-rays \citep{1991MNRAS.249..352G}. Fitting a fixed
reflection continuum (calculated from the code of
\citealp{1995MNRAS.273..837M}), with the reflector abundances tied to
the abundances of the plasma model, does not improve the best fit.
However, it does show that the data are consistent with a reflection
continuum with a total best fit $\chi^2_{\nu}$ of $1.3$ ($251$ d.o.f).

%% 7 spec - model = mgA parameters
%% uses P60005_7_nopcu0_backerr_0.5_time

\begin{figure}

\center{\includegraphics[width=0.9\columnwidth]{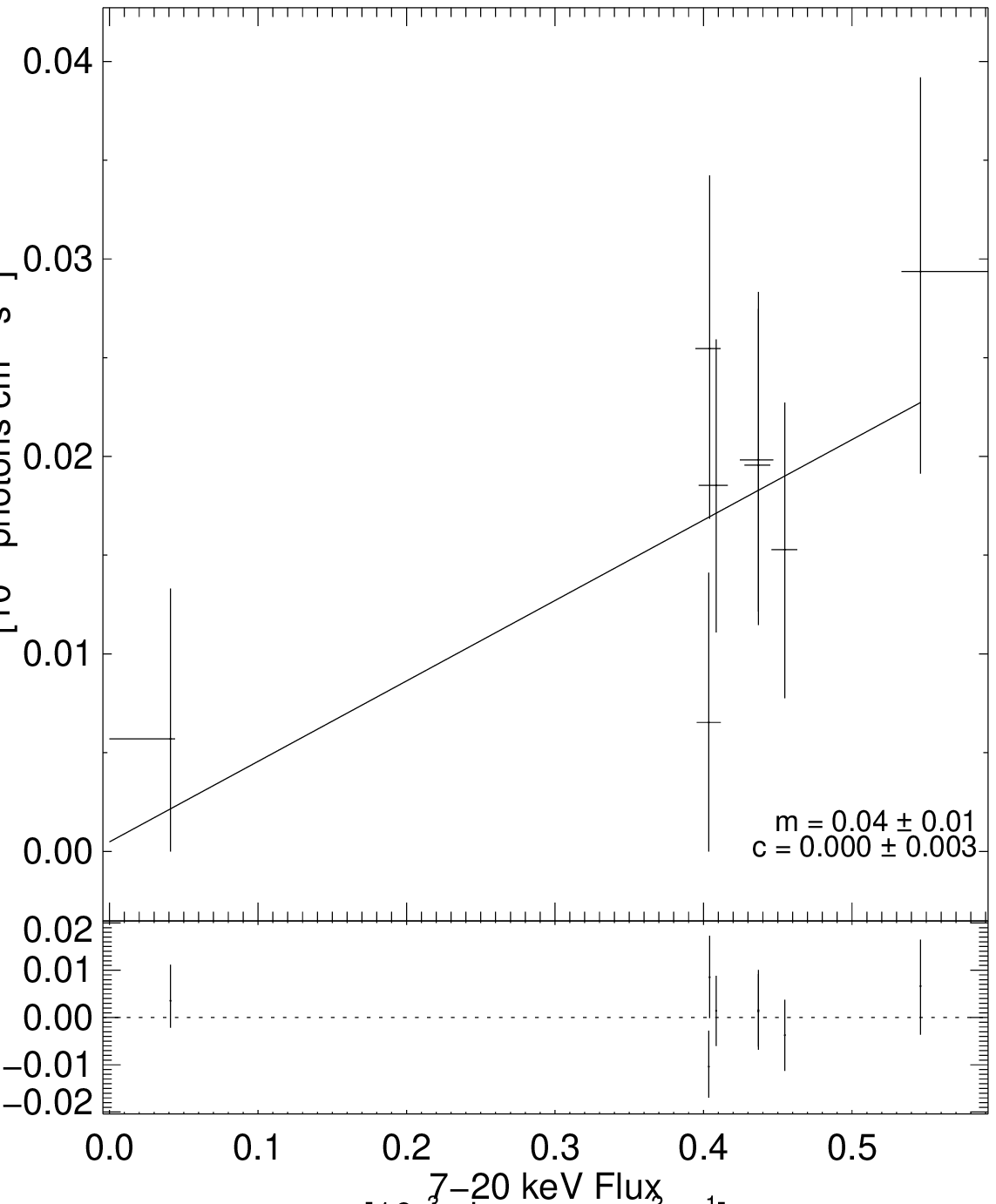}}
\vspace{0.3cm}
\caption{\label{pl_fluxline}The best fit $6.4$ keV line normalisation
    and the continuum flux above $7$ keV. Also plotted is the best fit
    line to the data points. The bottom panel shows the residuals
    between the best fit line and the data. All errors are in the $68$
    per cent confidence interval for one parameter of interest
    ($\Delta \chi^2 = 1.0$).} 
	
\end{figure}

%% line vs flux
It is possible to investigate the reflection in the system based on
the fluorescent line and flux above $7$ keV, since only photons above
$7$ keV have enough energy to be able to produce fluorescence photons
that contribute to the $6.4$ keV emission line. The line normalisation
and the line flux above $7$ keV are plotted in Fig. \ref{pl_fluxline}
which shows the fraction of continuum photons that produce line
photons.  If the amount of reflection in the system is constant then
as the continuum emission increases the line emission will also
increase. Thus a straight line is expected, with the gradient of the
best fit line in Fig. \ref{pl_fluxline} representative of the
reflection in the system.  As the flux from the X-ray source decreases
the number of photons producing fluorescence line photons also
decreases, thus the best fit line is expected to have an intercept
that passes near to zero. The figure shows there is a good correlation
with the flux between $7-20$ keV and the line normalisation. The
gradient of the line shows that $\sim 4$ per cent of hard X-rays give
rise to fluorescent photons.  The linear fit is consistent with no
change in reflection during these observations.

\subsection{Multi-temperature fits}

The X-ray spectrum originates from an optically thin plasma that
probably consists of a wide range of temperatures, previously modelled
as a cooling flow \citep{1996A&A...307..137W, 2005ApJ...626..396P}. We
attempted to fit a multi-temperature model to the combined quiescent
spectrum (cemekl in \xspec{}, \citealp{1996ApJ...456..766S}) which
showed that $\alpha \geq 0.8$, $T_{max} = 12^{+2}_{-4}$ keV and
abundance $= 0.59 \pm 0.05$ with a $\chi^2_{\nu}$ of $0.97$ ($33$ d.o.f)
confirming low abundances. $T_{max}$ and $\alpha$ are consistent with
\asca{} observations of other dwarf novae in quiescence
\citep{2005MNRAS.357..626B}.

\subsection{Fluxes and luminosity}
\label{sec_flux}

Fluxes were calculated from the best fitting model over the energy
range $2-20$ keV. Broad band fluxes were calculated by integrating
over the energy range $0.01-100$ keV. They are presented in Table
\ref{tab_suuma_mdot} along with the luminosities and associated
accretion rates. Assuming a distance of $260^{+190}_{-90}$ pc to
\suuma{} \citep{2003AJ....126.3017T} the average X-ray luminosity of
the system was calculated to be $2.6 \times 10^{32} (d/260pc)^2$ erg
s$^{-1}$ in quiescence and $0.7 \times 10^{32} (d/260pc)^2$ erg
s$^{-1}$ in outburst. We note that although the outburst flux is
consistent with that calculated by \cite{2005MNRAS.357..626B} it is
close to the confusion limit of the \rxte{} detector
\citep{2006ApJS..163..401J}. The accretion rates were calculated using
the relation $L=GM_{wd}\dot{M}/2R_{wd}$. The mass of the white dwarf
in \suuma{} has not been directly calculated owing to its low
inclination. Based on the work of \cite{2006MNRAS.373..484K} white
dwarfs in binaries were found to have an average mass of $M_{wd} =
0.75 M_{\odot}$ and $R_{wd} = 7.7 \times 10^8$ cm.  These values were
used to calculate the accretion rates presented here. The calculated
luminosities correspond to an average accretion rate during quiescence
of $4.2 \times 10^{15}$ g s$^{-1}$ and $1.2 \times 10^{15}$ g s$^{-1}$
during outburst.  However, the accretion rate is not thought to drop
during outburst, instead the luminosity is probably dominated by an
intense extreme ultraviolet component.

% insert observations table table_suuma_obs.tex
\begin{table}
  \begin{minipage}[c]{\columnwidth}

	\caption{\label{tab_suuma_mdot} Fluxes, luminosities and associated
		accretion rates for \suuma{}.}

%%
%% table with information about flux, luminosity and accretion rate
%% table includes:
%%		flux
%%		luminosity
%%		accretion rate
%%

%\begin{tabular*}{\textwidth}{p{3cm}@{\extracolsep{\fill}}cccc}
\begin{tabular*}{\textwidth}{l@{\extracolsep{\fill}}cccc}

\hline \hline

Time &
%$2-20$ keV Flux &
%$0.01-100$ keV Flux &
% \multicolumn{2}{c}{Flux [$10^{-11} $ ergs s$^{-1}$ cm$^{-2}$]} &
% \multicolumn{2}{c}{Flux\footnote{$\times 10^{-11} $ ergs s$^{-1}$ cm$^{-2}$}} &
% Luminosity\footnote{$\times 10 ^{32} (d/260pc)^2$ ergs s$^{-1}$} &
% Accretion rate\footnote{$\times 10^{15}$ g s$^{-1}$} \\

\multicolumn{2}{c}{Flux$^a$} &
Luminosity$^b$ &
Accretion rate$^c$ \\

[JD] &
$2-20$ keV &
$0.01-100$ keV &
& \\
% &[$(d/260pc)^2$ ergs s$^{-1}$] 
% &[$10^{15}$ g s$^{-1}$] \\

%&[$10^{-11} $ ergs s$^{-1}$ cm$^{-2}$] 
%&[$10^{-11} $ ergs s$^{-1}$ cm$^{-2}$] 
%&[$10^{32} (d/260pc)^2$ ergs s$^{-1}$] 
%&[$10^{15}$ g s$^{-1}$] \\

\hline

%% flux (2-20 0.01-100 keV)  luminosity  m_dot
2451999 	&1.7  &2.8  &2.3  &3.8 \\
2452013 	&1.9  &3.2  &2.6  &4.3 \\
2452028 	&1.8  &3.0  &2.4  &4.0 \\
2452042 	&1.8  &3.1  &2.5  &4.1 \\
2452055 	&1.9  &3.2  &2.6  &4.2 \\
2452068 	&1.8  &3.0  &2.4  &4.0 \\
2452079		&2.3  &3.9  &3.1  &5.2 \\
                                 
Outburst &0.4  &0.9  &0.7  &1.2 \\

\hline

\end{tabular*}

$^a$ $\times 10^{-11} $ ergs s$^{-1}$ cm$^{-2}$ \\
$^b$ $\times 10 ^{32} (d/260pc)^2$ ergs s$^{-1}$ \\
$^c$ $\times 10^{15}$ g s$^{-1}$ \\

  \end{minipage}
\end{table}

\section{Discussion and conclusions}
\label{Discussion}

We have presented X-ray observations of \suuma{} spanning six normal
outbursts.  We have studied the X-ray flux evolution of \suuma{} in
much greater detail than has been possible with the brief snapshot
observations that previous analysis has relied on.

%Outburst      	1.4472852
%Quiescence    	5.3116233

%% outbursts shapes
All six outbursts showed consistent X-ray behaviour.  At the start of
outburst the X-rays were suddenly quenched to near zero, the X-ray
count rate dropping by nearly a factor of $4$. This is consistent with
snapshot observations by \cite{1994ApJ...425..829S} who measured the
\rosat{} X-ray count rate to be a factor of $3$ lower in outburst.  A
larger drop in X-rays is measured from \rxte{} data since the spectrum
gets softer as well as fainter.  Observation of \vwhyi{}, also a
\suuma{} type variable, showed the flux to drop by a factor of $6$
\citep{1996A&A...307..137W}.

%% optical-X-ray delay
Calculating times for the optical and X-ray fluxes to cross the
mid-transition point indicated that the X-ray suppression lags behind
the optical rise by about half a day.  This is an indication of the time
the heating wave in the disc takes to propagate to the boundary layer.
%Different delay times observed for the separate outbursts might indicate
%that the heating wave originates from different parts of the disc
%requiring it to travel a different distance each outburst, although
%error on each individual delay is large, due to the typical cadence of
%the observations of $9-18$ hours. 
\sscyg{} has also been observed to have a delay between optical and
X-ray outbursts, with the beginning of X-ray outburst lagging behind
the optical by $0.9-1.4$ days \citep{2003MNRAS.345...49W}.  The
shorter delay for \suuma{} suggests that the time for the heating wave
to travel through the disc is shorter, perhaps due to its smaller
accretion disc.

%% increase at start of obs
The lack of an X-ray flux increase at the start of any of the
outbursts is puzzling since the boundary layer is only thought to
become optically thick once the accretion rate has reached a critical
value \citep{1979MNRAS.187..777P}.  In \sscyg{}, the X-rays do
increase before being suddenly suppressed \citep{2003MNRAS.345...49W}.
It may be that the quiescent accretion rate in \suuma{} is already
close to the critical rate.  However, our estimated accretion rate in
quiescence ($4.2 \times 10^{15}$ g s$^{-1}$) is below the expected
critical value of $2 \times 10^{16}$ g s$^{-1}$, although this does
depend on the white dwarf mass and radius, the viscosity of the disc
and temperature of the shocked gas \citep{1979MNRAS.187..777P}.  In
\sscyg{} the transition is seen at an accretion rate of $1 \times
10^{16}$ g s$^{-1}$ \citep{2003MNRAS.345...49W}.  Alternatively, the
observations presented here might have missed the flux increase.  The
optical rise in \sscyg{} was approximately $1.5$ days with an
associated duration of $12$ hours for the rise in the hard X-rays.
The optical rise in \suuma{} was approximately $1$ day but the average
separation of points in the composite light curve of Fig.
\ref{pl_xop_corr_suuma} results in an upper limit of $2$ hours to the
duration of a peak that occurs at the same outburst phase and with the
same duration in each case.  An increase in flux was also not seen in
\vwhyi{} \citep{1996A&A...307..137W}, although these observations had
a cadence of one observation a day.

The X-ray suppression occurred very rapidly and is not resolved in our
composite light curve, with a cadence of $2$ hours (Fig.
\ref{pl_q_declinei}). This is similar to \sscyg{} where the
suppression occurred in less than $3$ hours
\citep{2003MNRAS.345...49W}.

%% X-ray recovery
The X-ray recovery began while the optical band was in decline from
outburst.  This is earlier than seen in \vwhyi{} where the X-ray
recovery occurred at the end of the optical outburst
\citep{1996A&A...307..137W}.  Presumably this is due to a cooling
front reaching the boundary layer before passing through the whole
disc, perhaps suggesting that the cooling wave in \suuma{}  does not
start at the outer edge of the disc.

%% flux drops in Q
The X-ray count rates measured for \suuma{} tend to decrease during
quiescence, dropping by $12$ per cent over $8$ days. This is also seen
in \sscyg{} by \cite{2004ApJ...601.1100M}, where the decrease was by
$40$ per cent over $31$ days. It is interesting to note that in both
cases the count rate drops by about $1.7$ per cent day$^{-1}$, perhaps
indicating similar timescales acting in the inner accretion discs.
The decrease in X-ray flux is in direct conflict with the disc
instability model which predicts increasing quiescent accretion rates
\citep[e.g.][]{2001NewAR..45..449L}.

%% hardness -- vyhyi, yzcnc, wxhyi - suuma type
The X-ray spectrum of \suuma{} in outburst is softer and fainter than
during quiescence, in common with other dwarf novae
\cite[e.g.][]{2005MNRAS.357..626B}.  The X-rays observed during
quiescence arise from an optically thin region that is probably
replaced by an unseen optically thick emitting region that most likely
dominates during outburst.
%
%   T sscyg - Q - 20 keV OB - 5 keV \\
%   T suuma - Q -  8 keV OB - 4 keV \\
%   T  ugem - Q - 12 keV OB - 8 keV \\
%
Spectral fitting of the data shows that a thermal plasma model with
sub-solar abundances of $0.64 \pm 0.01$ and a $6.4$ keV line describes
the data well.  The data are also consistent with the presence of a
constant reflection continuum.

%% accretion rate
%
% L OB - 7.0x10^31
%    Q - 2.6x10^32
The luminosity during outburst, $7 \times 10^{31}$ ergs s$^{-1}$, was
similar to that calculated by \cite{2005MNRAS.357..626B} for \suuma,
with \asca{}. Other systems observed by \asca{} in outburst include
\rupeg{} ($9.4 \times 10^{31}$ ergs s$^{-1}$), a \ugem{} system.
During quiescence \suuma{} had an average luminosity of $2.6 \times
10^{32} (d/260pc)^2$ ergs s$^{-1}$, about as luminous as \sscyg{}
($3.6 \times 10^{32}$ ergs s$^{-1}$) but more luminous than \ugem{}
($2.8 \times 10^{31}$ ergs s$^{-1}$). \tleo{}, a \suuma{} system, was
also calculated to be less luminous with a luminosity of $1.8 \times
10^{31}$ ergs s$^{-1}$. The average luminosity during quiescence
corresponds to a quiescent accretion rate of $4.2 \times 10^{15}$ g
s$^{-1}$. This is similar to the quiescent rate of $3 \times 10^{15}$
found by \cite{2003MNRAS.345...49W} for \sscyg{} and is two and a half
orders of magnitude higher than predicted by the disc instability
model (e.g.  \citealp{2000A&A...353..244H}).

\section*{Acknowledgements}

We thank D. Baskill for his role in planning the observations
presented in this paper and the anonymous referee for constructive
comments. This research has made use of data obtained through the High
Energy Astrophysics Science Archive Research Centre Online Service,
provided by the NASA/Goddard Space Flight Centre.  We thank the
\aavso{} for providing optical light curves, which are based on
observations by variable-star observers worldwide.  DJC acknowledges
funding from an STFC Studentship.  Astrophysics research at the
University of Warwick is funded by an STFC rolling grant.

	\bibliography{refs}       %% Start your bibliography here; you can
	\bibliographystyle{mn2e}

\end{document}